\documentclass[preprint,prb,showpacs]{revtex4}
\usepackage{graphicx}
\usepackage{amsmath}
\usepackage{amsfonts}
\usepackage{color}

\begin{document}
\title{Single atom anisotropic magnetoresistance on a topological insulator surface}

\author{Awadhesh Narayan, Ivan Rungger, and Stefano Sanvito}
\affiliation{School of Physics, AMBER and CRANN Institute, Trinity College, Dublin 2, Ireland}

\date{\today}

\begin{abstract}
We demonstrate single atom anisotropic magnetoresistance on the surface of a topological insulator, 
arising from the interplay between the helical spin-momentum-locked surface electronic structure and 
the hybridization of the magnetic adatom states. Our first-principles quantum transport calculations based 
on density functional theory for Mn on Bi$_2$Se$_3$ elucidate the underlying mechanism. We complement 
our findings with a two dimensional model valid for both single adatoms and magnetic clusters, which leads 
to a proposed device setup for experimental realization. Our results provide an explanation for the conflicting
scattering experiments on magnetic adatoms on topological insulator surfaces, and reveal the real space 
spin texture around the magnetic impurity. 
\end{abstract}

\maketitle

\section{Introduction}
Topological insulators (TIs) are a materials class holding great promises for new avenues in both fundamental 
and applied condensed matter physics. In TIs a spin-orbit-driven bulk band inversion results in time-reversal 
symmetry-protected surface states. The spin-momentum locked surface states are robust against non-magnetic disorder 
and perfect backscattering is forbidden as a consequence of counter-propagating electrons possessing opposite spins, 
which cannot be flipped by time reversal symmetry obeying impurities~\cite{kane-rev,zhang-rev}. 

Conventional magnetoresistance (MR) devices, spin valves, utilize two magnetic electrodes (polarizer and analyzer) 
separated by a spacer. Recently, Burkov and Hawthorn found a new kind of MR on TI surfaces, which requires only 
one ferromagnetic electrode~\cite{hawthorn-mr}. TI surface-based spin valves, showing anomalous MR, have also 
been studied by model calculations~\cite{nagaosa-spinvalve}. 
In this paper we report an extremely large anisotropic single-atom MR on the surface of a 
TI, stemming from the interplay between the helical surface states and the spin anisotropy of the magnetic 
adatom. Based on this huge anisotropic MR (of several hundred percents, compared to a few percents 
usually obtained in conventional ferromagnets), we propose a new device concept, which has a number of 
advantages over previous proposals for magnetic sensors: (i) it does not need any magnetic electrode, but 
requires only a magnetic adatom, (ii) it does not rely on opening a band gap in the surface states, and (iii) it 
does not require injecting a spin polarized current into the topological insulator. 
Our idea is based on the magnetic anisotropy of atoms on TI surfaces, which theory predicts to be 
large~\cite{fazzio-adatoms} and magnetic circular dichroism 
confirms~\cite{wiesendanger-febise1,wiesendanger-febite}, demonstrating controllable magnetic 
doping~\cite{wiesendanger-febise2}. 

In presence of magnetic impurities spin-mixing on the surface of a TI is possible and scattering may be allowed. 
However, experiments have been conflicting and at the moment it is not clear whether or not scattering 
is observed~\cite{yazdani-dopant,madhavan-dopant}. Our results provide a possible way of reconciling these 
observations. We show that magnetic impurities open new backscattering channels, but these are found only at those 
energies where the impurity presents a large density of states and hybridizes with the underlying TI surface states. 
Then, the conductance depends strongly on the orientation of the local moment of the magnetic adatom, which implies 
a large MR. Away from these energies the transmission is close to the unperturbed value and no signature of the 
magnetic dopant is seen.  Our large-scale density functional theory (DFT) calculations allow us to probe the real space 
spin texture around the magnetic adatom, where the inclusion of atomistic details reveals significant differences from 
previous model-based calculations~\cite{zhang-magimp,balatsky-magimp}. 

\begin{figure}[h]
\begin{center}
  \includegraphics[scale=0.65]{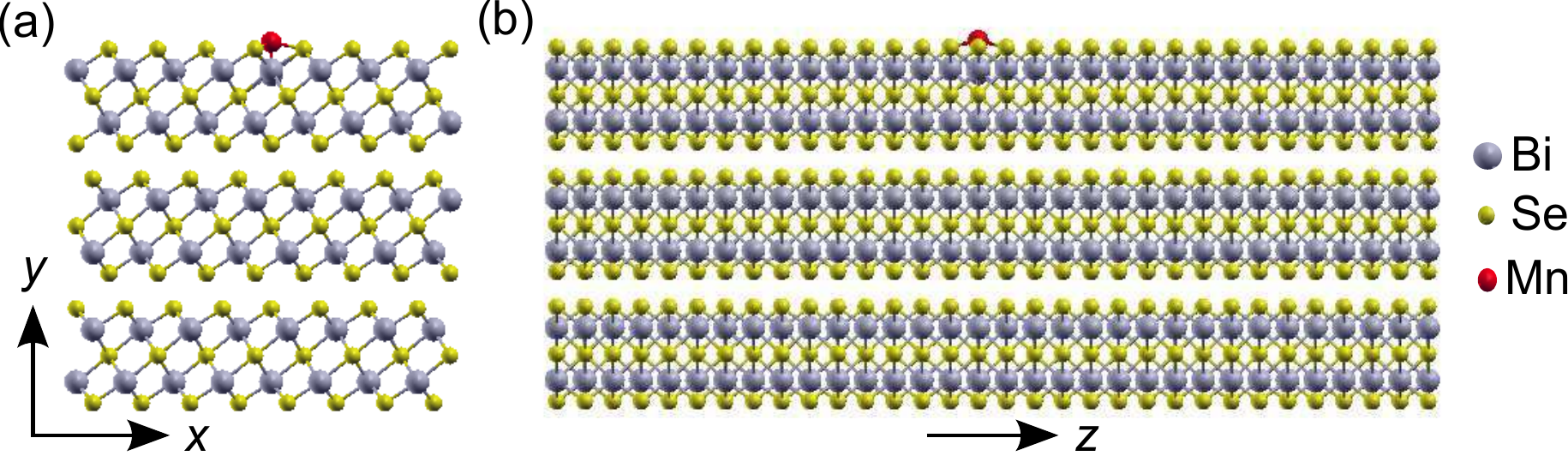}
  \caption{(Color online) Transport setup with a Mn atom adsorbed on a three quintuple layer Bi$_2$Se$_3$ slab, (a) viewed in the 
  plane perpendicular to and (b) along the transport direction ($z$). The scattering region supercell consists of 8 Bi$_2$Se$_3$ 
  primitive unit cells in the $xy$ plane and 16 unit cells along $z$, giving a concentration of 1 Mn atom in 1920 Bi/Se
  atoms ($\approx$ 0.05\%).}
  \label{setup}
  \end{center}
\end{figure}
%

\section{Computational Methods}
First-principles transport calculations are performed using the {\sc smeagol} code~\cite{sanvito-smeagol1,sanvito-smeagol2,sanvito-smeagol3}, 
which interfaces the DFT {\sc siesta} package~\cite{soler-siesta} to a non-equilibrium Green's function approach. Spin-orbit interaction is described 
by the on-site approximation~\cite{sanvito-soc}. An order-$N$ implementation allows us to study large systems with a few thousand atoms, while 
maintaining a good basis set quality~\cite{sanvito-sbqwell}. We use a double-$\zeta$ polarized basis, with a real space mesh cutoff of 300~Ry. 
The generalized gradient approximation for the exchange-correlation functional is used. The valence comprises Bi ($6s$, $6p$), Se ($4s$, $4p$) 
and Mn ($3d$, $4s$), while norm-conserving Troullier-Martins pseudopotentials describe the core electrons. Mn is studied in 
the Bi on-top geometry, which is the most stable~\cite{fazzio-adatoms}. All the atoms in the top quintuple layer (QL) are allowed to move and the 
structures are relaxed until the forces are less that 0.001 eV/\AA. For transport calculations, semi-infinite electrodes comprising 3 QL Bi$_2$Se$_3$
slabs are attached to the scattering region (shown in Fig.~\ref{setup}), and a minimum of 25 \AA{} of vacuum is included along the slab thickness 
($y$-direction). We use a $3\times 1\times 1$ $k$-point grid for converging the charge density, while a much denser grid of at least 80 $k_{x}$-points 
is employed to evaluate the transmission, reflection amplitudes and densities of states.  

\section{Results and Discussion}
\begin{figure}[t]
\begin{center}
  \includegraphics[scale=0.65]{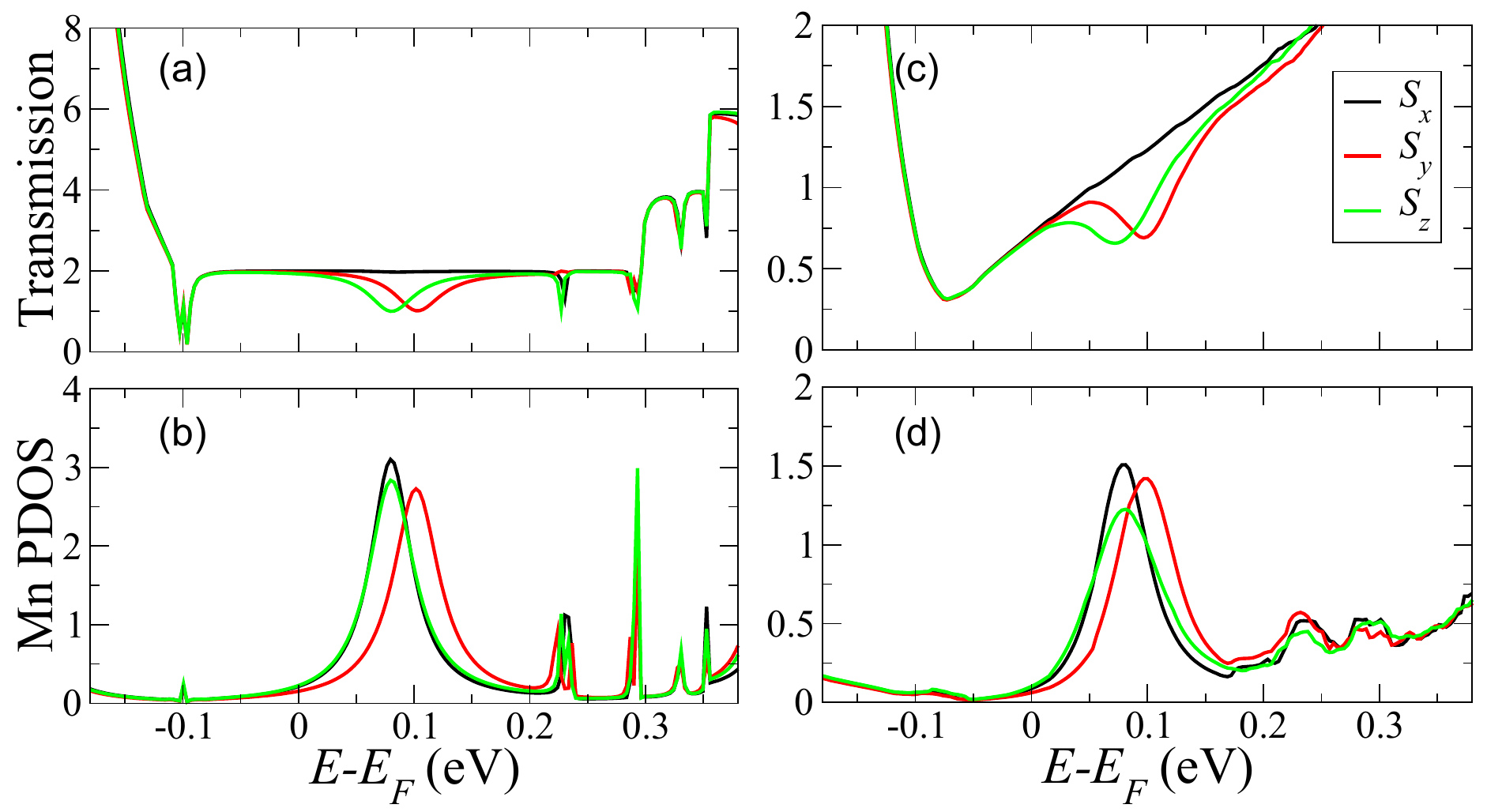}
  \caption{(Color online) Transmission and density of states projected on Mn for different Mn spin directions. 
  In (a) and (b) results are for $k_{x}=0$, while in (c) and (d) they are averaged over all the incidence angles. 
  For Mn spin along $x$, the transmission is unperturbed, while reduced transmission occurs for other directions, 
  resulting in a single-atom anisotropic MR.}
  \label{trms}
\end{center}
\end{figure}

The transmission coefficient for different orientations of the Mn magnetic moment is shown in Fig.~\ref{trms}. 
For $k_{x}=0$ [normal incidence, Fig.~\ref{trms}(a)] and the Mn spin aligned along the $x$-axis, which is  
the spin direction of the incoming electron's traveling along the positive $z$-axis, the transmission is close to
two in the energy window of the topological state (approximately -0.1~eV to 0.3~eV), i.e. there is a unity contribution 
from each surface. In contrast, for the two orthogonal Mn spin directions, a dip in transmission occurs in the energy 
range, where a peak in the Mn projected density of states (PDOS) is found. When the Mn spin is along $x$, there is 
no reduction in transmission, even though there is a peak in Mn PDOS with a height comparable to the case of 
the other two directions. In all the three cases we find the Mn adatom having a moment close to 4.5~$\mu_\mathrm{B}$, 
in agreement with previous reports~\cite{fazzio-adatoms}, and a substantial in-plane magnetic anisotropy of 
6~meV~\cite{anisotropy}. After integrating over the entire Brillouin zone for all $k_{x}$ values a similar 
picture is obtained [see Figs.~\ref{trms}(c) and \ref{trms}(d)]. Thus, we find that at the energies of the Mn states 
there emerges an \emph{anisotropic} MR, depending upon the spin orientation of the magnetic adatom. We find a 
MR, $\mathrm{MR}=(T_{x,s}-T_{\alpha,s})/T_{\alpha,s}$, of 670\% (here $T_{x,s}$ is the transmission at the top surface 
with the Mn spin along $x$ and $T_{\alpha,s}$ is the surface transmission for the other two Mn spin directions $\alpha=y, z$). 

\begin{figure}[tb]
\begin{center}
  \includegraphics[scale=0.65]{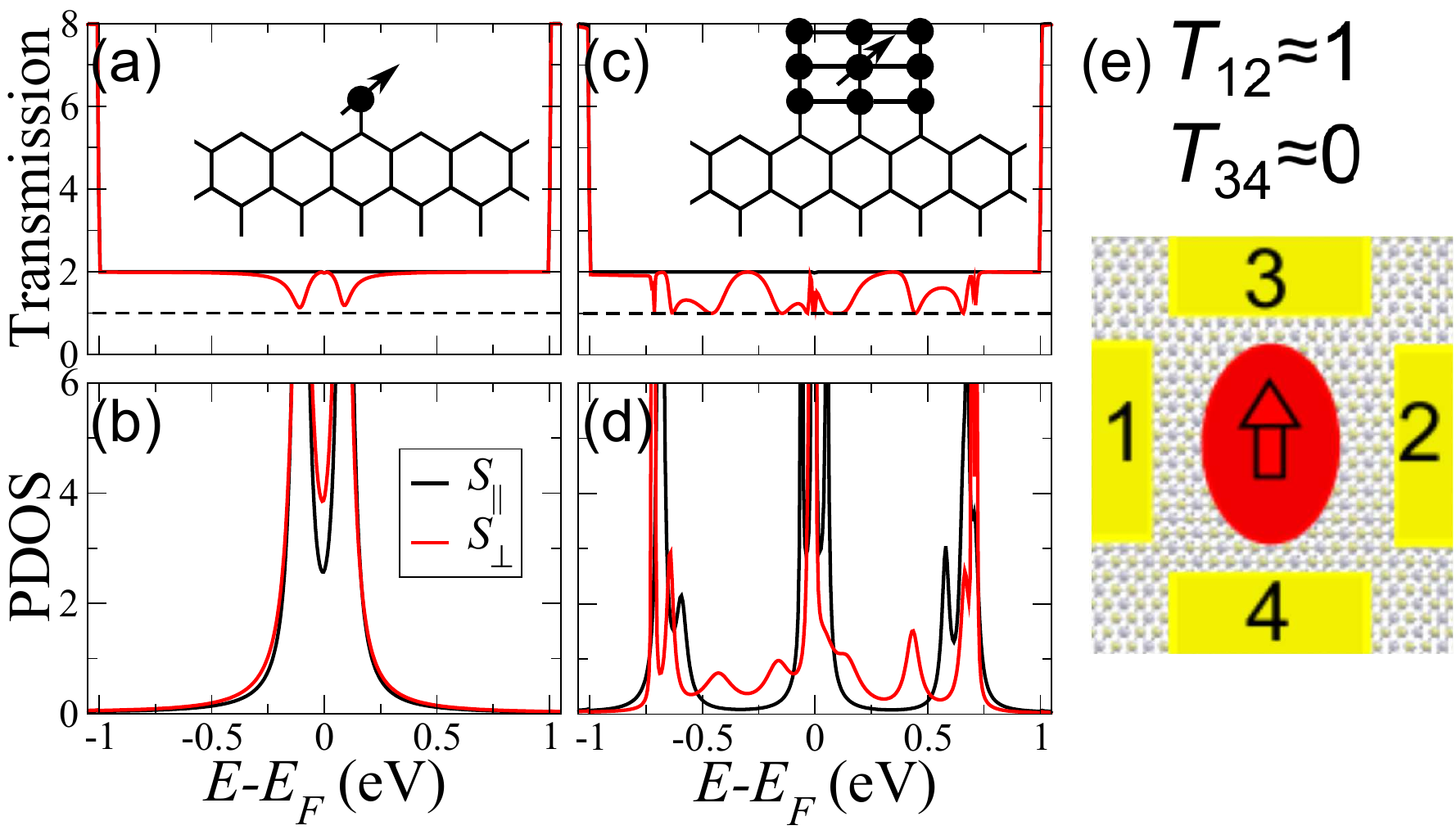}
  \caption{(Color online) (a) Transmission and (b) adatom PDOS for the two-dimensional model, with the adatom spin 
  pointing parallel and perpendicular to the electron spin. (c) Transmission and (d) average PDOS for a magnetic cluster 
  in the two spin configurations. The insets are schematic of the two setups and the dashed lines 
  indicate the transmission of one from the unperturbed edge. Here the adatom onsite energy is 0.1, the hopping 
  integrals to the ribbon 0.3, the hopping between magnetic atoms 0.5 (in units of the nearest neighbor hopping) 
  and the other parameters are the same as in Ref.~[\onlinecite{sanvito-andreev}]. (e) Schematic of the four-probe 
  geometry proposed to measure the anisotropic MR.}
  \label{trms-km}
\end{center}
\end{figure}

We emphasize that this mechanism for MR does \emph{not} involve opening a band gap in the surface 
state spectrum. Our findings can be compared to the results in Ref.~\cite{rauch-dual}, where, in the presence of 
a magnetic field, the gap in the surface state spectrum does not open as long as the mirror symmetry is preserved. 
This is a consequence of the dual topological character of chalcogenides, which are strong topological insulators as 
well as topological crystalline insulators. This mirror symmetry protects the Dirac crossing, however, the Dirac crossing 
can be shifted away from the time reversal invariant momenta. Applying a magnetic field perpendicular to a mirror plane 
of the crystal lattice breaks time-reversal symmetry and destroys the ${Z}_{2}$ topological phase, while the topological 
crystalline phase is still present, although with the Dirac point shifted away from $\Gamma$. In this case since we have 
a mismatch between the states from the electrodes (where the Dirac point has not been shifted off $\Gamma$) and the 
Dirac states close to the magnetic impurity, we find a high resistance state.
In this particular setup the transport is along the $z$ direction and, for normal incidence, 
the spin-momentum relation locks the spin of the surface state along $x$. If the Mn impurity 
spin points along this direction, then electrons suffer minimal scattering and the resistance is low, while 
for other Mn spin directions we find a high resistance state. In contrast, if the electrodes are positioned in 
the orthogonal configuration, such that transport is along $x$, then the propagating electron spin will be along 
$z$. In this case the low resistance state will be obtained for the Mn spin parallel to the direction of propagation,
$z$, while the other two directions will yield a high resistance state [see Fig.~\ref{trms-km}(e)]. Since 
the resistance is given by the orientation of the local magnetic moment with respect to the transport direction, 
this MR is also anisotropic. In case of strong hexagonal warping, for instance in Bi$_2$Te$_3$, there is also a finite out-of-plane spin component~\cite{henk-warping}. For transport along the $x$-direction, the impurity magnetization along $z$ would not be 
parallel to the spin of the propagating electron and this would result in a high resisting state.

\begin{figure}[b]
\begin{center}
  \includegraphics[scale=1.0]{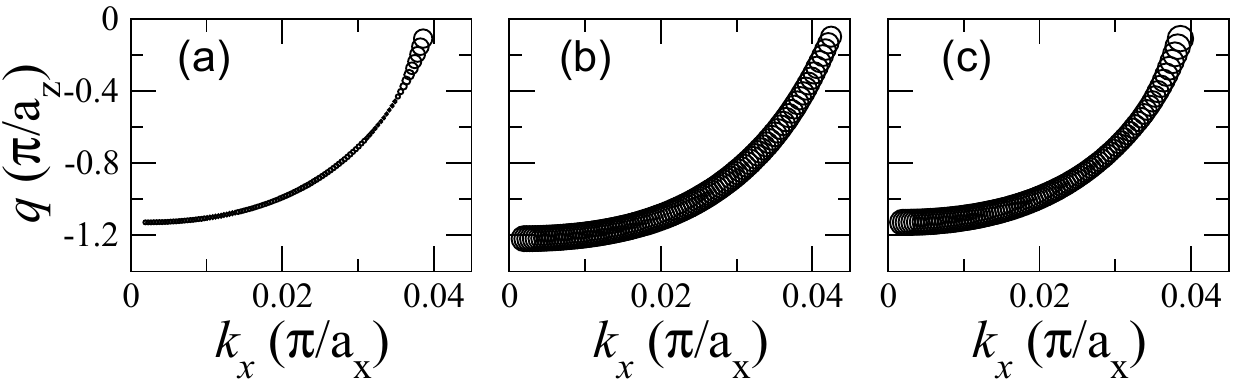}
  \caption{Scattering vectors, $q$, as a function of the incident wave vector, $k_{x}$, for Mn spin along 
  (a) $x$, (b) $y$, and (c) $z$. The size of the circles is proportional to the reflection amplitude. 
  The curves are plotted at energies corresponding to peaks in Mn density of states, $E-E_\mathrm{F}$= 0.08~eV, 
  0.10~eV and 0.08~eV, for Mn spin along $x$, $y$ and $z$, respectively. Here $a_{x}$ and $a_{z}$ are lengths of 
  the electrode unit cell along $x$ and $z$ directions.} 
  \label{scatvec}
\end{center}
\end{figure}

From the previous results it is not possible to unequivocally distinguish whether the scattering occurs due to 
spin-flip between states on one surface or if the MR is an artifact of inter-surface scattering caused by the finite 
Bi$_2$Se$_3$ slab thickness. We clarify this issue by calculating the full scattering matrix and evaluating 
the transmission and reflection amplitudes for the individual scattering states on the top and bottom 
surfaces~\cite{sanvito-bisestep2}. We obtain inter-surface reflection and transmission amplitudes always smaller 
than 0.008. For intra-surface scattering, in contrast, these quantities reach values up to 1, which confirms that the slab 
is thick enough to prevent significant coupling between opposite surfaces. 

A deeper analysis is provided by studying the scattering wave vectors, $q$, and the reflection amplitudes, $r$, on the 
top surface of the TI slab at the peak energy in Mn PDOS, as a function of the wave vector $k_{x}$ along the direction 
perpendicular to transport. Here $q=k_{z,\mathrm{out}}-k_{z,\mathrm{in}}$ is the difference between the outgoing, 
$k_{z,\mathrm{out}}$ and incoming, $k_{z,\mathrm{in}}$, $z$-components of the scattering wave vectors. Since in 
the bulk gap both $k_{z,\mathrm{in}}$ and $k_{z,\mathrm{out}}$ for the topological surface states are functions of 
$k_x$, we can evaluate $q$ as function of $k_x$. The result is in Fig.~\ref{scatvec}, with the size of the circles 
denoting the reflection amplitude $r(k_{x})$. Since the constant energy surface in the energy range of the topological 
states is approximately circular, we also find the corresponding $q$-$k_x$ plot having a circular shape.  
For the Mn spin along $x$ and for small $k_x$ the reflection amplitude is vanishingly small, while it becomes larger 
when $k_x$ increases. This is because the overlap between the two counter-propagating surface states 
get larger when $k_{x}$ increases. Thus, when the Mn spin is along $x$ the impurity behaves as a non-magnetic 
scattering center~\cite{sanvito-bisestep1}. In contrast, for the other two directions a large reflection is present even for 
$k_{x}=0$, which persists at larger $k_{x}$. The total reflection is obtained by integrating this function over all $k_x$, so 
that the underlying difference in reflection amplitude for small $k_x$ is what yields the anisotropic MR.

Recent scanning tunneling microscopy studies of magnetic adatoms on TI surfaces have observed either new scattering 
channels, to be ascribed to magnetic scattering~\cite{madhavan-dopant}, or found the scattering independent of the magnetic 
nature of the adatom~\cite{yazdani-dopant}. Our calculations provide a possible explanation for these conflicting observations. In the case of non-magnetic impurities there is scattering, but only for $k_{x}\neq 0$. Additional scattering, which can also occur at $k_{x}=0$, is found for certain directions of the moments of the magnetic impurities. However, this happens only at the energies where the impurity atoms present a large density of states. The transmission coefficients show that a new backscattering channel is created only at the energy of the adatom PDOS, while at all other energies where the topological state exists, no fingerprint of the magnetic adatom is visible. Thus, a likely explanation to reconcile experiments is that the adatom should not only hybridize with the TI surface, but also present 
peaks in density of states at relevant energies for being detected in the transmission spectra. These depend on the specific magnetic atom and the adsorption site and therefore can differ in different experiments. Away from these energies the transmission spectrum resembles the case of non-magnetic impurities.

\begin{figure*}[ht]
\begin{center}
  \includegraphics[scale=0.65]{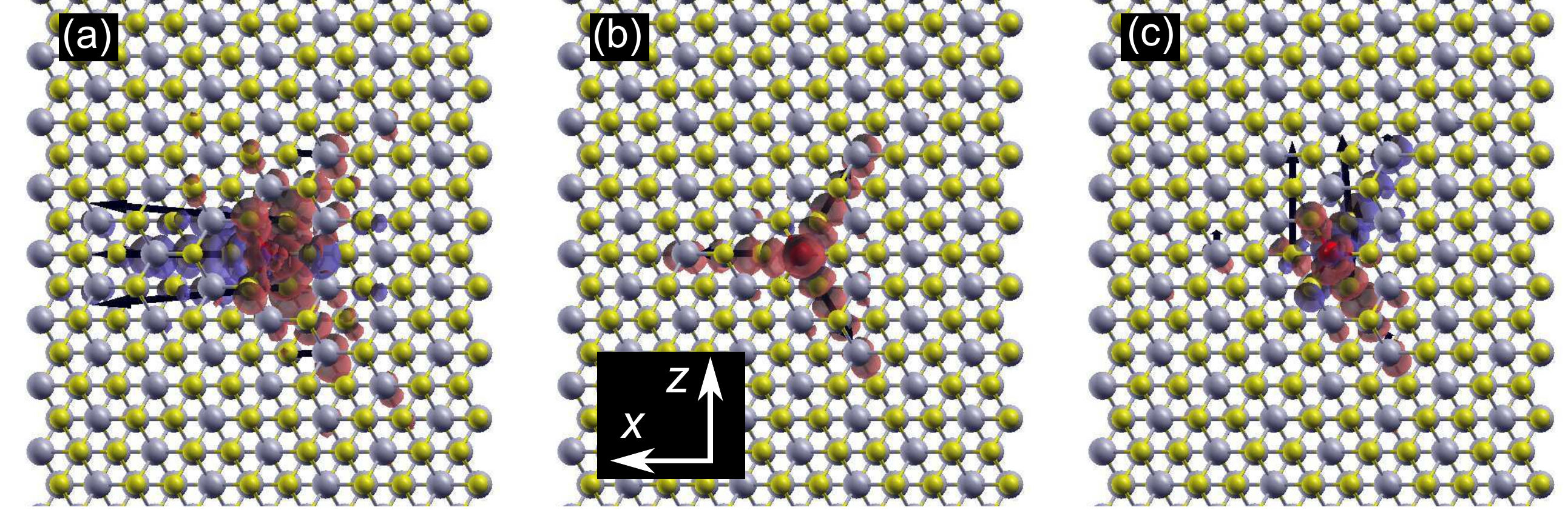}
  \caption{(Color online) Local DOS at the energy of the peak in Mn DOS, showing the real space spin texture 
  around the magnetic adatom with its spin pointing along (a) $x$, (b) $y$, and (c) $z$. The arrows denote the 
  in-plane spin components obtained from the atom-projected DOS. The isosurfaces correspond to the local DOS 
  projected along the direction normal to the plane, with red denoting positive values and blue negative. The 
  effect of the adatom spin in not limited to the top surface Se atoms, but it is distributed over the first QL.}
  \label{spintex}
\end{center}
\end{figure*}

The anisotropic MR can be understood by considering the impurity as the source of an effective local magnetic field. 
If the spin of the adatom is parallel to the spin of the propagating electron such an effective field provides a
collinear scattering potential, thus precluding spin mixing and backscattering. However, if the local spin forms an 
angle with that of the itinerant electrons, opposite spin electrons will couple and thus backscattering between helical 
states will become possible. A minimal two-dimensional model can be used to verify the generality of the MR. We 
use the Kane-Mele model~\cite{kanemele} for a ribbon with a magnetic adatom or a magnetic cluster placed at the 
ribbon edge and an exchange coupling between the electron spin and the impurity~\cite{sanvito-andreev}. The edge 
electrons in this model capture the essential physics of the $k_{x}=0$ case of three-dimensional TIs, which is responsible 
for the anisotropic MR. The results are shown in Fig.~\ref{trms-km}. The transmission is high for the adatom spin parallel 
to the electron spin, while it is low for other angles, thus that the model calculations confirm our first-principles results. 
Furthermore, for the magnetic cluster the MR is obtained over an energy range larger than that of the single adatom. 
The fact that the anisotropic MR is independent of the details of adatom means that one can select other magnetic ions 
to tailor the anisotropy direction. For instance, Cr and Co on Bi$_2$Se$_3$ exhibit an out of plane easy axis, while Mn 
and Fe an in plane one~\cite{fazzio-adatoms}. 

In addition to a two-terminal device the anisotropic MR can be measured in a four-probe setup [Fig.~\ref{trms-km}(e)]. When 
the impurity spin points in the direction shown (e.g. due to the magnetic shape anisotropy), then a measurement of the 
resistance between the electrodes $1$ and $2$ yields a low resistance state, while high resistance is measured between 
$3$ and $4$. If a thin film with in-plane magnetization is used, then an MR will be obtained depending on the in plane 
orientation of the magnetization. Out-of-plane magnetization, in contrast, always yields a high resistance state. 
In general, when the impurity spin points parallel to the helical electron spin the resistance is low, while other angles between 
the two spins will result in a higher resistance. A large magnetic anisotropy also implies the likely absence of Kondo-type 
features, which occur with degenerate ground states. Furthermore we expect the spin-flip of the impurity to be negligible 
as long as the bias is smaller than the magnetic anisotropy~\cite{sanvito-iets,sanvito-iets2}. Going a step further, we have found the 
same anisotropic MR for magnetic clusters, in which the aforementioned effects will be even smaller and the magnetic 
anisotropy may be engineered to be large.

Since our \textit{ab initio} calculations employ extremely large supercells, we are in the position to probe the real-space 
spin texture around the isolated magnetic impurity. This has been previously studied with Dirac-like effective 
Hamiltonians~\cite{zhang-magimp,balatsky-magimp}, but here the full details of the electronic structure are included. 
A combination of atom projected DOS and local DOS is shown in Fig.~\ref{spintex} for the three different orientations 
of the Mn spin at the energy corresponding to the peak in Mn PDOS of any given orientation. The induced spins on 
the atoms around Mn are predominantly along the direction of the Mn spin. For Mn spin pointing along $y$, we find 
a hedgehog-like in-plane spin texture, with the spins pointing outwards from the impurity site. This contrasts continuum 
models, which yield a vortex-like in-plane structure~\cite{zhang-magimp,balatsky-magimp}. The out-of-plane spin points 
along the positive $y$ direction, in agreement with the model results. This spin is induced over the first QL. For Mn spin 
along $y$, the spin texture exhibits a three-fold rotational symmetry of the underlying lattice, which is not captured by the 
continuum low-energy model. For the other two directions, this lattice symmetry is broken by the Mn spin and the neighboring 
atoms exhibit a spin along the impurity spin direction. We have also investigated the spin texture at other energies and found 
similar directions as those presented in Fig.~\ref{spintex}, although the magnitude of the induced spin decreases at energies 
away from Mn PDOS peak. Our spin texture predictions naturally call for an experimental corroboration via spin-polarized 
scanning tunneling microscopy~\cite{heinrich-stm1,heinrich-stm2}.

\section{Summary and Conclusions}
In conclusion we have discovered single-atom anisotropic magnetoresistance on topological insulator surfaces 
decorated with magnetic adatoms. This effect is a consequence of the spin-momentum locking of TI surface states 
interacting with the adatom spin. The MR does not originate from the opening of a gap in the surface band structure, 
nor from spin injection. Furthermore, our results provide a possible explanation for the conflicting observations 
concerning scattering from magnetic atoms on TI surfaces. Our order-$N$ code allowed us to study the real space 
spin texture around the adatom, which has differences from previous model calculations. Based on these findings 
we propose magnetoresistive devices with potentially large MR, utilizing either single magnetic atoms or thin film nanodots 
incorporated between non-magnetic electrodes, using an in plane rotation of the thin film magnetic moment.    

\section*{Acknowledgments}
AN thanks the Irish Research Council for financial support (EMBARK initiative). 
IR and SS acknowledge support from KAUST (ACRAB project). We thank Aaron Hurley and Andrea Droghetti 
for useful discussions and Alin Elena for technical support at ICHEC. Computational resources have been 
provided by Irish Centre for High-End Computing (ICHEC) and Trinity Centre for High Performance Computing.


\begin{thebibliography}{99}
\bibitem{anisotropy} We have also verified this value by plane wave calculations using {\sc vasp} code for a smaller $3\times 1\times 3$ supercell.
\end{thebibliography}
\end{document}